\documentclass[%
reprint,
superscriptaddress,
nofootinbib,
amsmath,amssymb,
aps,
prb,
floatfix,
longbibliography
]{revtex4-2}

\usepackage{graphicx}
\usepackage{bm}
\usepackage{xcolor}

\usepackage[mathlines]{lineno}

\usepackage[ignoreunlbld,norefs,nocites]{refcheck}
\usepackage[colorlinks=true, linkcolor=blue,urlcolor=blue,citecolor=blue]{hyperref}
\usepackage{cleveref}

\begin{document}

\preprint{APS/123-QED}

\title{Adiabatic Ramp Dynamics Across the ETH--MBL Transition in Disordered XXZ Spin Chain}

\author{Nidhi Kumari}
\affiliation{%
Computational Quantum Many-Body Physics Lab, Department of Physics, Dr.\ B.\ R.\ Ambedkar National Institute of Technology, Jalandhar, Punjab - 144008, India}%

\author{Vinod Ashokan}
\email{ashokanv@nitj.ac.in}
\affiliation{Computational Quantum Many-Body Physics Lab, Department of Physics, Dr.\ B.\ R.\ Ambedkar National Institute of Technology, Jalandhar, Punjab - 144008, India}

\date{\today}
\begin{abstract}

Many-body localization(MBL) provides a mechanism by which isolated interacting quantum systems with disorder can avoid thermalization unlike ergodic systems satisfying the eigenstate thermalization hypothesis(ETH). Many-body localized systems retain signatures of their initial conditions at long times, whereas systems obeying ETH lose such information as they approach thermal equilibrium. Studying Non-equilibrium dynamics across ETH-MBL crossover is an important problem in condensed matter physics. Adiabatic control of parameters in interacting disordered systems provides a powerful framework to investigate MBL phases and their dynamical robustness. Using exact diagonalization and time-dependent numerical methods we study the effects of adiabatically ramped interactions in a disordered spin-1/2 XXZ chain, a paradigmatic model for exploring the many-body localization transition. By monitoring diagonal entropy density and entanglement entropy density growth across various ramp speeds, and system sizes. Our study incorporates Finite-size effects of spectral observables to probe the transition between ergodic and localized phases. The numerical results show that localized dynamical behavior remains largely intact under sufficiently slow ramp evolution, while increasing the driving rate promotes stronger excitation generation and larger entropy growth. This trend highlights the strong dependence of nonequilibrium adiabatic dynamics in disordered interacting quantum many-body systems.

\end{abstract}

 \maketitle


\section{Introduction}
The emergence of a thermal balance in isolated quantum many body systems remains a central point in non equilibrium quantum statistical mechanics \cite{Deutsch1991,Srednicki1994,Rigol2008}. Because such systems evolve unitarily and they are not coupled to some outside environment, the origin of thermal behavior is  not immediately obvious. The eigenstate thermalization hypothesis provides a widely accepted explanation, by saying that each eigenstate of a nonintegrable Hamiltonian already carries thermal type properties. So, the average values of local observables end up matching with what we have obtained from thermal ensembles at the same energy density, and then isolated systems can effectively thermalize by themselves through internal dynamics only. In a sense, this picture connects thermalization to ergodicity and also quantum chaos. The theoretical foundations of ETH were developed from the pioneering works of Deutsch and Srednicki, who basically claimed that interactions cause particles to group together into complex many-body eigenstates so that statistical behavior can show up already at the eigenstate level. Later, extensive numerical investigations strengthened the whole picture. In particular, studies on quenched interacting lattice, systems suggested that generic nonintegrable models do relax toward thermal equilibrium during unitary time evolution, whereas integrable setups that harbor extensive conservation laws develop long time behavior that is qualitatively different, and they are instead explained by generalized statistical ensembles. These developments established ETH as the default framework for thermalization in generic interacting quantum systems, and it also became useful reference point to identify when thermalization fails to happen or breaks down in closed quantum systems \cite{Deutsch1991,Srednicki1994,Rigol2008}.
A basically different dynamical regime shows up in strongly disordered interacting setups, where many-body localization can prevent thermalization . If the absence of interactions, strong disorder leads to Anderson localization and then inhibits transport. But whether that kind of localization survives against interactions was a, long-standing question because many-body system is usually assumed to promote ergodicity. Early theoretical studies suggested that disorder and interactions can hold up a stable nonergodic state.

Early theoretical work suggested that localization can survive even when interactions are there in disordered quantum systems. In particular, Gornyi \textit{et al.} \cite{Gornyi2005} argued that interaction effects do not always destroy localization, while Basko, Aleiner, and Altshuler \cite{Basko2006} then developed a more microscopic framework to pred the existence of a localized phase at nonzero energy density. After that, many numerical studies offered a strong support for many-body localization. When people study the spectral properties, they saw a change from Wigner–Dyson level statistics, which is the fingerprint of ergodic behavior, to Poisson statistics, that is what we expect for localization \cite{Oganesyan2007}. When we study the disordered spin chains using simulations we see a crossover between thermalizing and localized regimes \cite{Pal2010,Nandkishore2015,Altman2015,Vosk_2013}. A commonly used theoretical description of the localized phase based on an extensive set of quasi local conserved quantities called as local integrals of motion ($\ell$-bits), and it is the idea that supports the robustness of many-body localization.

The dynamics of the entanglement is an another important probe for localization. After giving a quantum quench, ergodic systems end up building entanglement fater, while many-body localized state shows a much slower growth of the entanglement entropy, it follows an almost logrithmic behaviour \cite{Bardarson2012,Bar_Lev_2014}. After some time there are experiments based on ultracold atoms, trapped ions, and other programmable quantum devices \cite{Geraedts2016,Khemani2017,Ovadia2015,Serbyn2015,Smith2016}. These experiments shows a precise control on disorder strength and interaction parameters, and driving protocols with high precision, and it follows the nonequilibrium time evolution. One important example for this experiment is ultracold fermions in quasiperiodic optical lattices, where the density imbalance stayed for a long time, and the system kept a meemory of the initial state for e very long time, so this gives a solid evidence for many-body localization \cite{Smith2016,Schreiber2015,Bordia2016}.
Furthur, subsequent experimetal studies provide a clarificiation about the role of the interactions, dissipation and disorder strength. The trapped-ion quantum simulator experiment confirms the signature of thermalization and exposed slow relaxation in disordered spin models. All of these experimental results shifts the theoretical phenomena of  many-body localzation into an experimental phenomena which we can actually see in experiments.Consequently a lot of research was done to figure out what is the MBL phase and how we can crossover to this new phase from ergodic phase. Because of these developments, how disordered interacting systems respond under time dependent driving attracts a lot of attention. While parameter changes that are slow enough are usually supposed to produce adiabatic evolution, trying to get a strict adiabatic limit in intracting disorderd many-body systems is not very easy. To get this we face trouble bacause of the interactions, avoided level crossings and finite-size effects, and there is complicated spectral texture due the diorder in the sytem, all of this make the system to not follow adiabaticity. We study this for slow quenches, periodically driven systems and more. In interacting disordered many-body system, the back and forth between the localization and external drive creates a nonequilibrium effects, that look different what from what people see the thermalizing system. Because transport gets suppressed , dephasing proceeds slowly, and memory of the initial conditions lasts for a very long time, localized phases end up responding to time-dependent perturbations in a fundamentally different manner. Earlier studies has already mentioned unusual adiabatic properties for localized systems, including nonlocal responses to a local perturbations and strongly restricted energy absorption under when the system is externally driven.These findings indicate that the dynamics near the ETH--MBL crossover are controlled not just by disorder and interactions but also by how exactly the driving protocol is implemented.
Here we look at the effect of different driving protocols on the adiabatic dynamics of a disordered spin-$1/2$ XXZ chain \cite{Luitz2015}. We use exact diagonalization and real-time numerical evolution method. First we re-obtain known behaviors for linear ramps, then we extend the study to test quadratic and exponential driving protocols. The resulting dynamics we obtain are diagnosed using entanglement entropy density, and diagonal entropy density, for system sizes up to $L=14$. Finally, by contrasting the response between ergodic and localized regimes under various ramp profiles, we evaluate how the driving protocol influences excitation production, entropy generation, and the quality of adiabatic state preparation in disordered interacting quantum systems.
The disordered XXZ model and the driving protocols considered in this work are introduced in Section \ref{Theoretical model}. The numerical framework with the observables are outlined in Section \ref{Numerical methods}. Then in Section \ref{Results} we discuss the results for the dynamical response across the ETH--MBL crossover, and we compare linear, quadratic, and exponential driving protocols using  entanglement entropy density plus diagonal entropy density. The final conclusions are presented in Section \ref{conclusion}.
\section{Theoretical Model and Formalism}
\label{Theoretical model}
We study the nonequilibrium dynamics of a disordered spin-$\frac{1}{2}$ XXZ chain, given by

\begin{equation}
	H(t)=
	\sum_{j=0}^{L-2}
	\left[
	\frac{J_{xy}}{2}
	\left(
	S_{j+1}^{+}S_j^{-}
	+\mathrm{h.c.}
	\right)
	+
	J_{zz}(t)S_{j+1}^{z}S_j^{z}
	\right]
	+
	\sum_{j=0}^{L-1}
	h_j S_j^{z},
	\label{eq:Hamiltonian}
\end{equation}

with $S_j^{\pm}=S_j^x\pm iS_j^y$ and $S_j^z$ is the usual spin-$\frac{1}{2}$ operator. The onsite disorder field terms $h_j$ are chosen independently, from a sample of uniform distribution in the interval $[-h_0,h_0]$. In all the computations, we keep the exchange constant fixed at $J_{xy}=1$ , but the interaction piece $J_{zz}(t)$ is tuned according to the driving scheme described right after. To separate the dynamical regimes cleanly we take two values, namely $h_0=0.1$ and $h_0=3.72$. The first one is meant to yield delocalized many-body motion, and the second is intended for localized behavior \cite{Znidaric2007,Luitz_2017}.

\subsection{Driving Ramp Protocols}
The nonequilibrium evolution is generated by introducing a time dependence in the interaction parameter $J_{zz}(t)$, while the rest of the Hamiltonian parameters stay unchanged. To examine the effects of the different driving protocols on the dynamical properties of the syatem. At a glance we see few standard ramp protocols, linear, quadratic, and exponential.

The linear ramp protocol is defined as

\begin{equation}
	J_{zz}^{\mathrm{lin}}(t)=J_{zz}(0)
	\left(
	\frac{1}{2}+vt
	\right),
	\label{eq:linear_ramp}
\end{equation}

in which $v$ is the ramp speed, and we fix $J_{zz}(0)=1$ as the energy unit. The ramp evolution during $0\leq t\leq t_f$, where

\begin{equation}
	t_f=\frac{1}{2v},
\end{equation}.

For comparison, we also defined a quadratic ramp protocol as

\begin{equation}
	J_{zz}^{\mathrm{quad}}(t)=J_{zz}(0)
	\left(
	\frac{1}{2}+vt^2
	\right),
	\label{eq:quadratic_ramp}
\end{equation}

Where the corresponding time span is,

\begin{equation}
	t_f=
	\sqrt{\frac{0.5}{v}},
\end{equation}

Lastly, an exponential ramp protocol is defined as 
\begin{equation}
	J_{zz}^{\mathrm{exp}}(t)=J_{zz}(0)
	\left(
	\frac{1}{2}e^{vt}
	\right),
	\label{eq:exponential_ramp}
\end{equation}

and here the corresponding time span is,

\begin{equation}
	t_f=
	\frac{\ln 2}{v}.
\end{equation}

These ramp protocols trace different trajectories in parameter space, enable a systematic comparison finite-rate driving between ergodic, and localized regimes. By tuning the ramp geometry but keeping both the initial and final interaction strengths fixed, we can study how the protocol itself affects excitation creation, entanglement increase, and breakdown of adiabatic evolution \cite{Rigol2008beyond,Rigolin2010,DeGrandi2010,Pollmann_2013,Kolodrubetz_2017,Maraga_2016,Weinberg2017,Serbyn_2014,Khemani_2015,Khemani_2016,Ni2023}.

\section{Numerical Methods}
\label{Numerical methods}

We use exact diagonalization to study the driven dynamics of disordered XXZ chains with open boundary conditions, and system sizes up to $L=14$. Computations are limited to the zero-magnetization sector,

\begin{equation}
	\sum_{j=0}^{L-1} S_j^z = 0.
\end{equation}

For each realization of disorder, the starting state is chosen as the eigenstate that lies closest to the centre of the spectrum of the initial Hamiltonian. The disorder-averaged quantities are obtained from roughly $10^3$ disorder samples for $L=10$ and $12, $and about $10^2$ disorder samples for $L=14$. Everything is performed with the QuSpin exact-diagonalization setup \cite{Weinbergqu2017,Weinberg2019}.

\subsection{Diagonal Entropy Density}

To quantify the extent of nonadiabatic excitations generated during the ramp, we compute the diagonal entropy density,

\begin{equation}
	s_d=\frac{S_d}{L},
	S_d=-\sum_n p_n \ln p_n,
	\qquad
\end{equation}

where the diagonal ensemble is defined as
\begin{equation}
	\rho_d=\sum_n p_n |n\rangle\langle n|,
	\qquad
	p_n=|\langle n|\psi(t_f)\rangle|^2,
\end{equation}

and here $|n\rangle$ labels the eigenstates of the final Hamiltonian $H(t_f)$.  The quantity $S_d$ corresponds to the Shannon entropy associated with the occupation probabilities that appear in the diagonal ensemble and it tells us how spread out the time-evolved state becomes when we look at the eigenbasis of $H(t_f)$.

If the protocol were perfectly adiabatic, the final state remains confined to a single eigenstate and $s_d$ tends to zero. Deviations from this ideal limit lead to a nonzero diagonal entropy, which is essentially a sign for the redistribution of spectral weight over many-body eigenstates during the evolution. Consequently, $s_d$ provides a useful measure of excitation generation under finite-rate driving \cite{Rigol2008beyond,DeGrandi2010,Weinberg2017,Polkovnikov2011}.

\subsection{Entanglement Entropy Density}

To see how quantum correlations end up building during the time evolution, we evaluate the bipartite entanglement entropy density,
\begin{equation}
	s_{\mathrm{ent}}=
	\frac{S_A}{|A|},
\end{equation}
where
\begin{equation}
	S_A=-\mathrm{Tr}
	\left(
	\rho_A \ln \rho_A
	\right)
\end{equation}

is the von Neumann entropy of the reduced density matrix

\begin{equation}
	\rho_A
	=
	\mathrm{Tr}_{A^c}
	\left(
	|\psi(t_f)\rangle
	\langle\psi(t_f)|
	\right).
\end{equation}

In this setup the subsystem $A$ is basically one half of the chain, here $|A|=L/2$, and $A^c$ is its complement . The value $s_{\mathrm{ent}}$ provides a measure for the correlations created during the ramp , and serves as a independent diagnostic for the dynamical features of the driven system.

We get the entanglement entropy density directly from the final state $|\psi(t_f)\rangle$ , and after that we average it over different disorder realizations \cite{Luitz_2017,Khemani_2015,Ni2023,Polkovnikov2011}.

\section{Results}
\label{Results}

\subsection{Spectral signatures of the ETH--MBL crossover}

Before studying the nonequilibrium dynamics, we first characterize the static spectral properties of the disordered XXZ chain. To this end, we first study the disorder-averaged adjacent-gap ratio, a widely used indicator of spectral correlations in interacting disordered systems. For consecutive many-body energy levels, the adjacent-gap ratio is defined as

\begin{equation}
	r_n=\frac{\min(\delta_n,\delta_{n+1})}{\max(\delta_n,\delta_{n+1})},
\end{equation}

where $\delta_n = E_{n+1}-E_n$ denotes the spacing between neighboring energy eigenvalues. Averaging $r_n$ over the spectrum and disorder realizations yields the quantity $r$.
In the weak-disorder regime, the many-body spectrum exhibits strong level repulsion and follows the statistical properties of the Gaussian orthogonal ensemble (GOE), characterized by an average gap ratio $r_{\mathrm{GOE}}\approx0.53$. In contrast, sufficiently strong disorder suppresses spectral correlations, resulting in Poisson-like level statistics with $r_{\mathrm{P}}\approx0.39$. The evolution of $r$ between these two limiting values therefore serves as a useful diagnostic of the crossover from ergodic to localized behavior.

\begin{figure}[!t]
	\centering
	\vspace{-0.4cm} 
	\includegraphics[width=\linewidth]{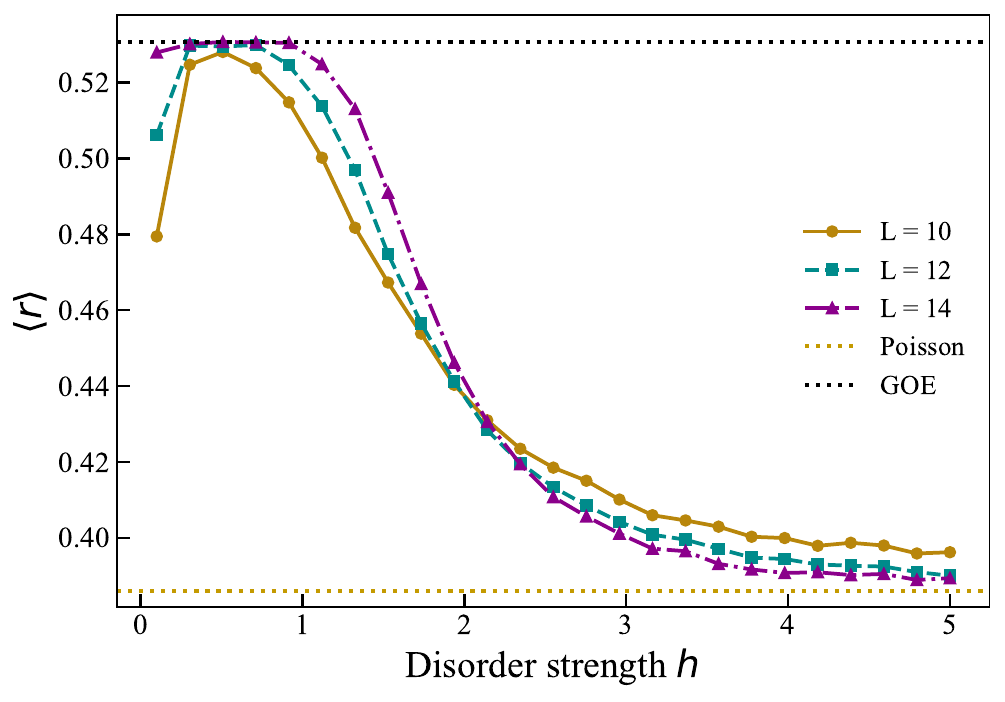}
	\caption{Disorder-averaged adjacent-gap ratio $r$ versus disorder strength for system sizes $L=10,12,$ and $14$.}
	\vspace{-0.3cm} 
	\label{fig:1}
\end{figure}

Figure~\ref{fig:1} shows the disorder-averaged ratio $r$ of adjacent-gap versus disorder strength for system sizes, $L=10,12, and 14$. In the weak-disorder regime, $r$ stays pretty close to the GOE value, consistent with ergodic dynamics and strong level repulsion. As disorder is increasing, $r$ goes down toward the Poisson limit, which points to a gradual quenching of spectral correlations, and then the many-body localization starts to show up. The crossover also becomes more clear when the system size increases, indicating reduced finite-size effects, and a sharper distinction between the ergodic and localized regimes. Even though exact diagonalization limits us to $L\leq14$, the observed trends are systematic across all accessible sizes, and they provide a reliable identification of the disorder regime associated with the ETH--MBL crossover. This spectral characterization establishes the parameter range is explored in the rest of this work, where we investigate the system's response to linear, quadratic, and exponential ramp protocols through the behavior of the diagonal entropy density, entanglement entropy density, and their finite-size scaling properties. The dashed horizontal lines indicates the Gaussian orthogonal ensemble (GOE) value $r_{\mathrm{GOE}} \approx 0.53$, which confirms ergodic behavior With increasing $h_{\text{disorder}}$, $r$ decreases monotonically toward the Poisson value $r_{\mathrm{P}} \approx 0.39$, reflecting the emergence of many-body localization. The gradual crossover from GOE to Poisson statistics with increasing disorder strength signals the transition, from the thermal phase, to the many body localized phase.

\subsection{Ramp-Driven Entropy Dynamics across the ETH--MBL Crossover}

\begin{figure*}
	\centering
	\begin{minipage}[t]{0.48\textwidth}
		\includegraphics[width=\textwidth]{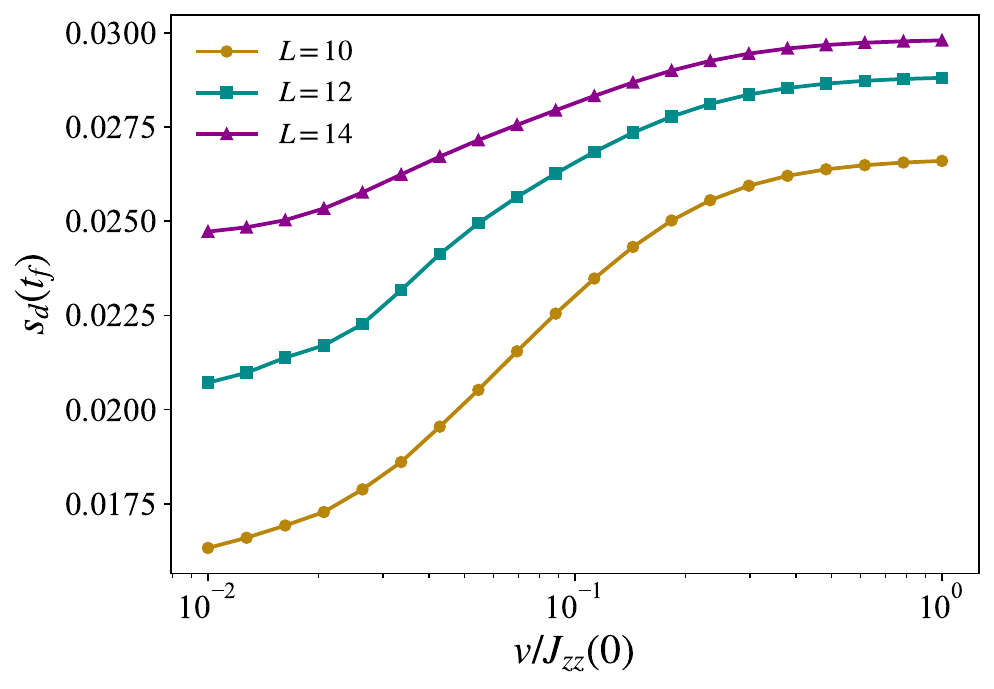}
	\end{minipage}
	\hfill
	\begin{minipage}[t]{0.48\textwidth}
		\includegraphics[width=\textwidth]{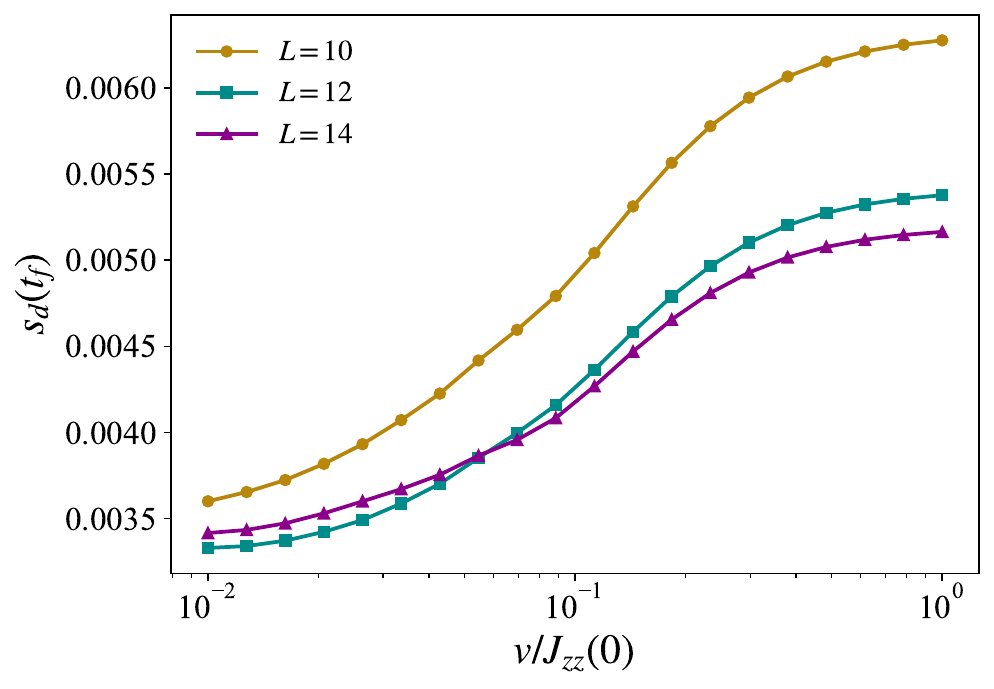}
	\end{minipage}
	
	\vspace{0.4cm}
	\begin{minipage}[t]{0.48\textwidth}
		\includegraphics[width=\textwidth]{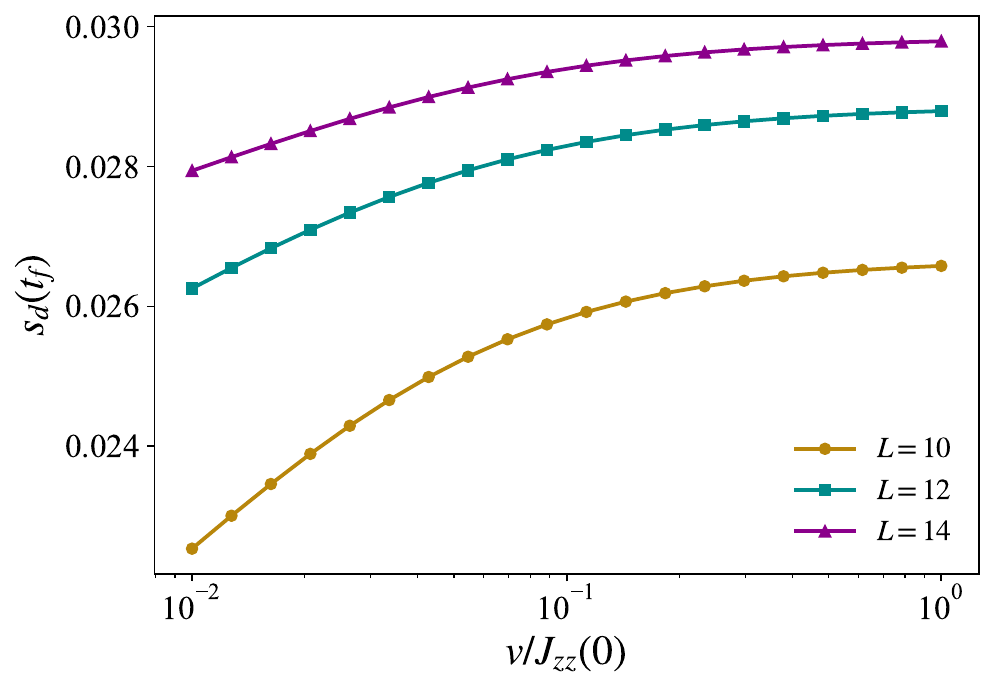}
	\end{minipage}
	\hfill
	\begin{minipage}[t]{0.48\textwidth}
		\includegraphics[width=\textwidth]{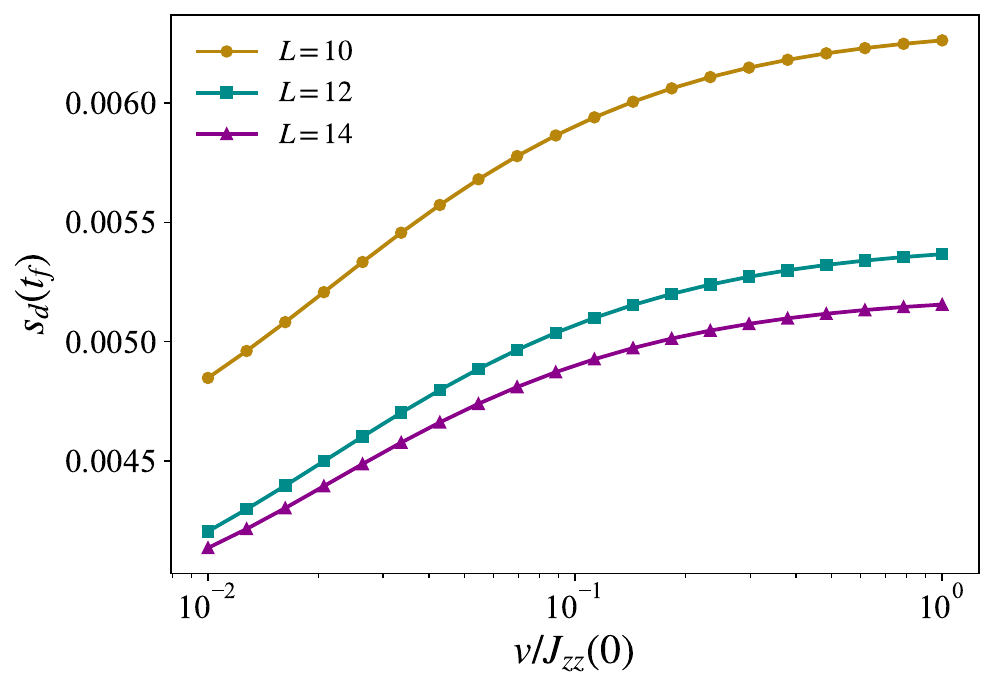}
	\end{minipage}
	
	\vspace{0.4cm}
	\begin{minipage}[t]{0.48\textwidth}
		\includegraphics[width=\textwidth]{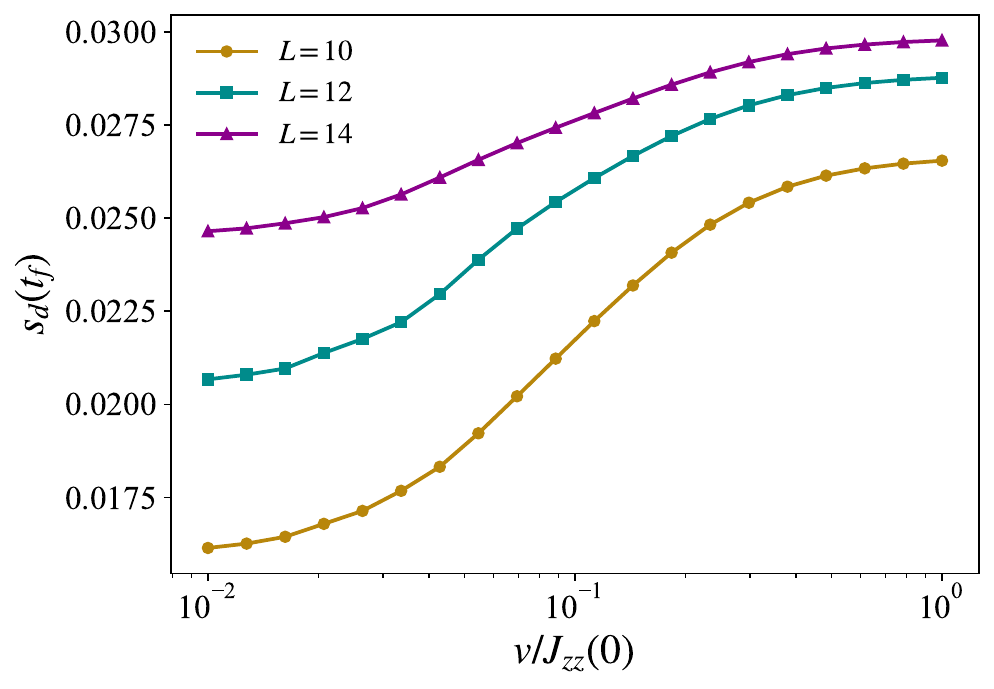}
	\end{minipage}
	\hfill
	\begin{minipage}[t]{0.48\textwidth}
		\includegraphics[width=\textwidth]{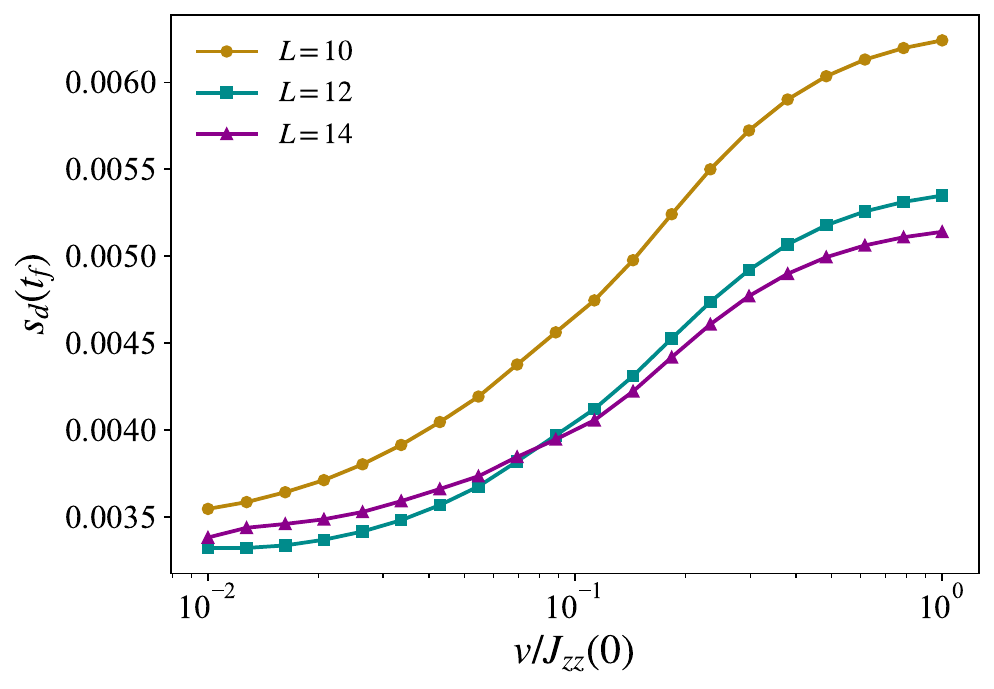}
	\end{minipage}
	
	\caption{Dependence of the diagonal entropy density, $s_d(t_f)$, on the normalized ramp rate, $v/J_{zz}(0)$, for the disordered XXZ spin chain. Panels in the left and right columns correspond to the ETH and MBL regimes, respectively. The rows represent linear, quadratic, and exponential ramp protocols from top to bottom. Results are shown for system sizes $L=10$, $12$, and $14$ to illustrate finite-size dependence and the influence of different driving protocols on entropy generation during nonequilibrium dynamics.}
	\label{fig:2}
\end{figure*}

\begin{figure*}
	\centering
	\begin{minipage}[t]{0.48\textwidth}
		\includegraphics[width=\textwidth]{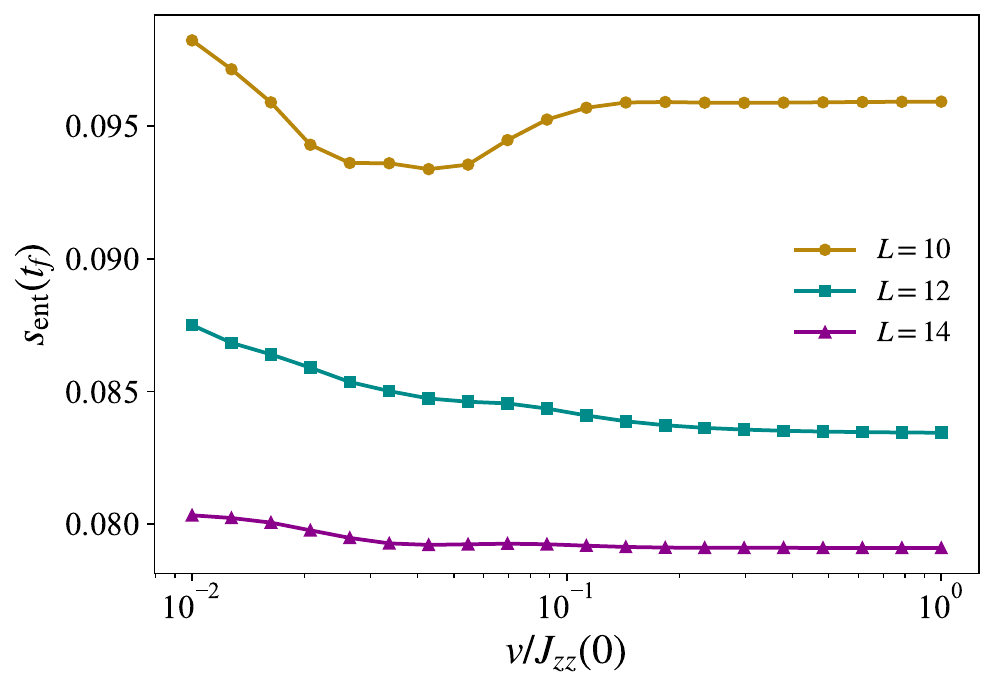}
	\end{minipage}
	\hfill
	\begin{minipage}[t]{0.48\textwidth}
		\includegraphics[width=\textwidth]{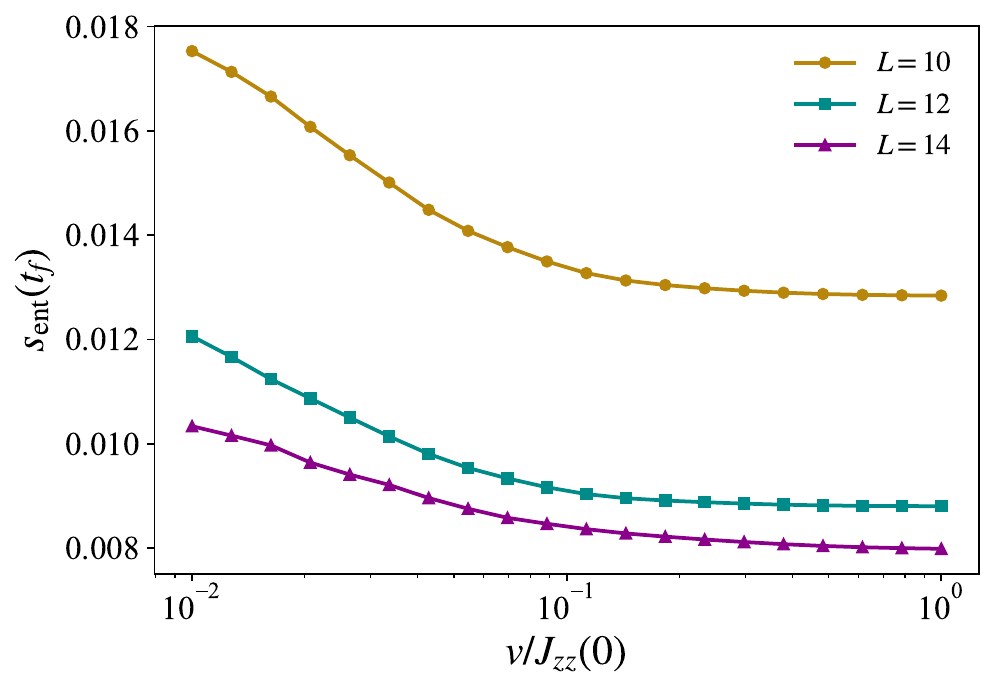}
	\end{minipage}
	
	\vspace{0.4cm}
	\begin{minipage}[t]{0.48\textwidth}
		\includegraphics[width=\textwidth]{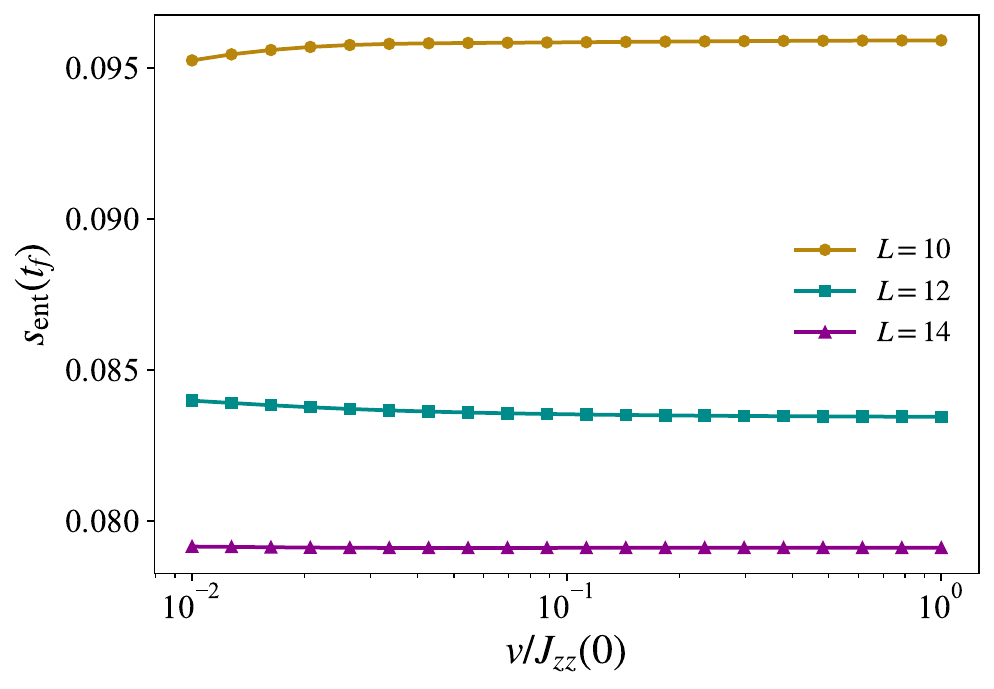}
	\end{minipage}
	\hfill
	\begin{minipage}[t]{0.48\textwidth}
		\includegraphics[width=\textwidth]{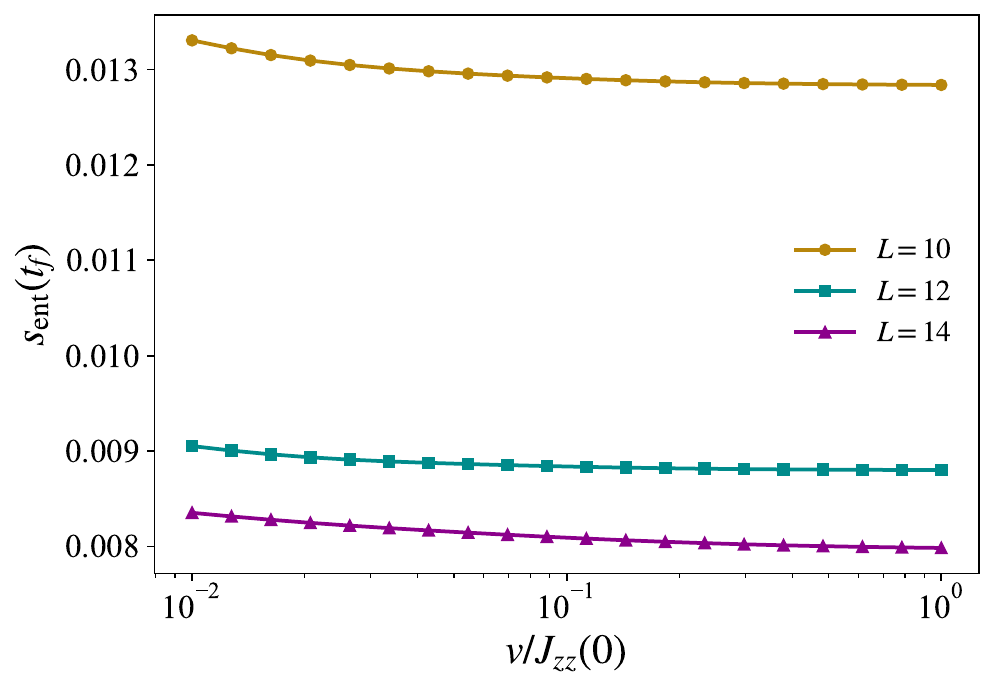}
	\end{minipage}
	
	\vspace{0.4cm}
	\begin{minipage}[t]{0.48\textwidth}
		\includegraphics[width=\textwidth]{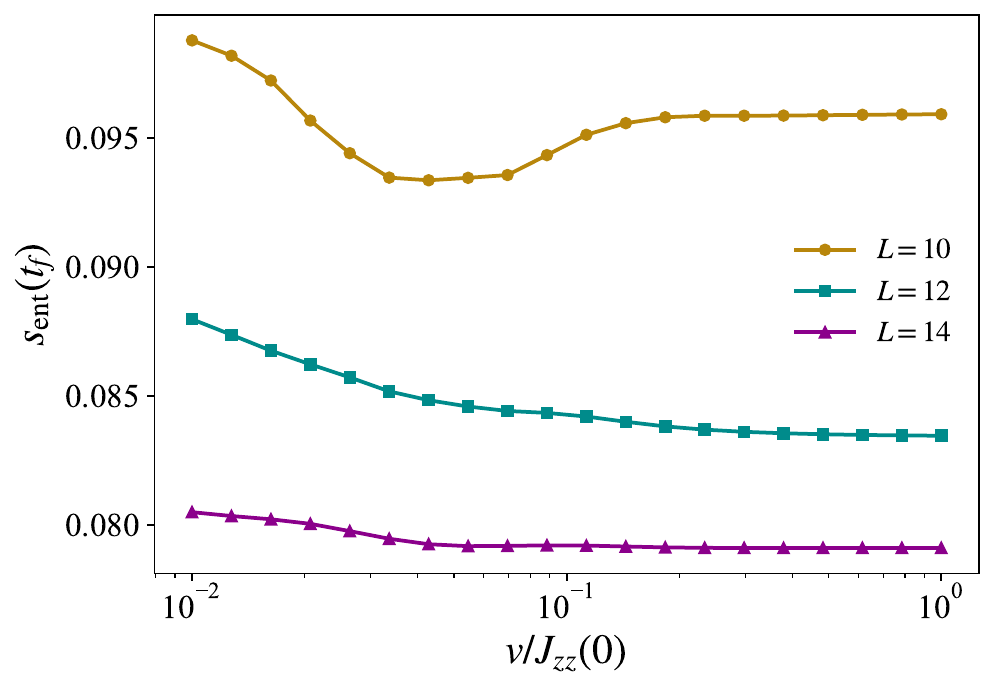}
	\end{minipage}
	\hfill
	\begin{minipage}[t]{0.48\textwidth}
		\includegraphics[width=\textwidth]{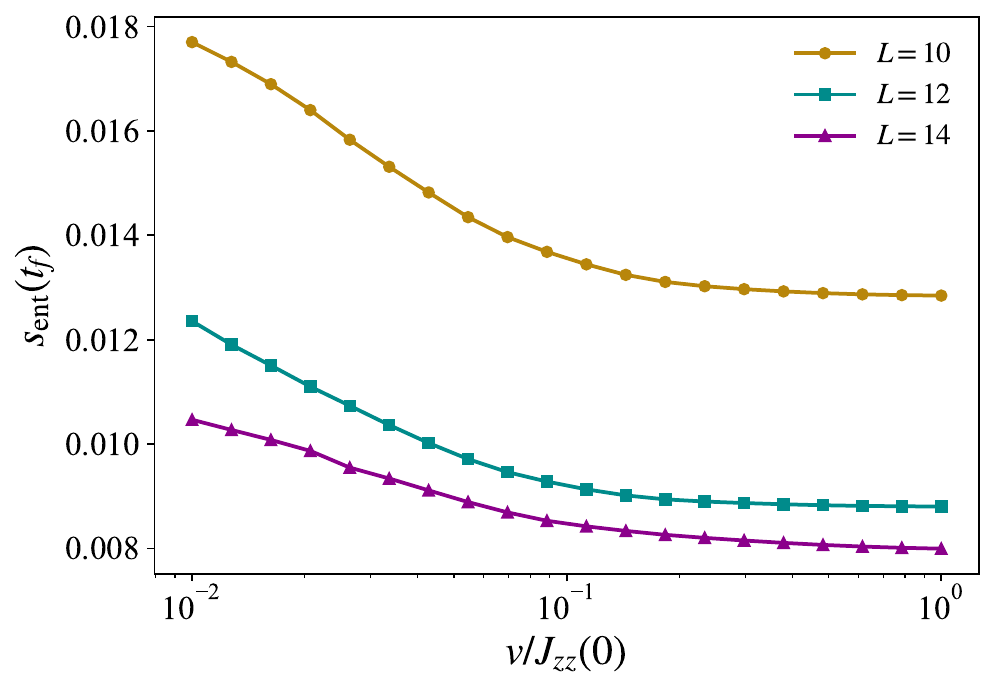}
	\end{minipage}
	
	\caption{Behavior of the entanglement entropy density, $s_{\mathrm{ent}}(t_f)$, with normalized ramp rate, $v/J_{zz}(0)$, in the disordered XXZ chain. Left and right columns show results in the ETH and MBL regimes, respectively, while the rows correspond to linear, quadratic, and exponential driving protocols. Data for system sizes $L=10$, $12$, and $14$ are included to examine how system size and ramp protocol affect entanglement growth generation during nonequilibrium evolution.}
	\label{fig:3}
\end{figure*}

Figures~\ref{fig:2} and \ref{fig:3} show how the diagonal entropy density $s_d(t_f)$ and the entanglement entropy density $s_{\mathrm{ent}}(t_f)$ change, as a function of the normalized ramp speed $v/J_{zz}(0)$, for linear , quadratic , and exponential driving protocols. The data are shown for both the ETH and MBL regimes, and we consider the system sizes $L=10$, $12$, and $14$. What stands out is that there is a clear contrast between the two phases, for every ramp protocol we tried. In the ETH regime, both entropy measures increase with ramp speed, indicating strong departures from adiabatic evolution, and also enhanced generation of many-body correlations. This increase becomes a bit more noticeable when the system size increases, which matches the idea that the larger Hilbert space gives the ergodic dynamics more space.
In that localized regime, Both $s_d(t_f)$ and $s_{\mathrm{ent}}(t_f)$ stay rather small all the way across the investigated range of ramp speeds, and they only show minor finite-size variations. The weak response  basically tells us that only a limited participation of many-body states really participates during the evolution, and they basically limit both the excitation creation and the correlation spreading, even while the system is driven. While every driving protocol keeps the same qualitative kind of dynamical behavior, still there are noticeable quantitative differences. The linear ramp produces the largest change in both  entropy measures, but the quadratic one ends up giving a more confined variation. The exponential ramp looks like the smoothest dependence on ramp speed, and it shows reduced fluctuations. Despite these differences, the overall scaling behavior remains largely unchanged. Taking it together, these results support diagonal entropy and entanglement entropy as complementary probes for the ETH–MBL crossover. Their behavior show a robust separation between ergodic and localized dynamics.

\begin{figure*}
	\centering
	
	\begin{minipage}[t]{0.48\textwidth}
		\includegraphics[width=\textwidth]{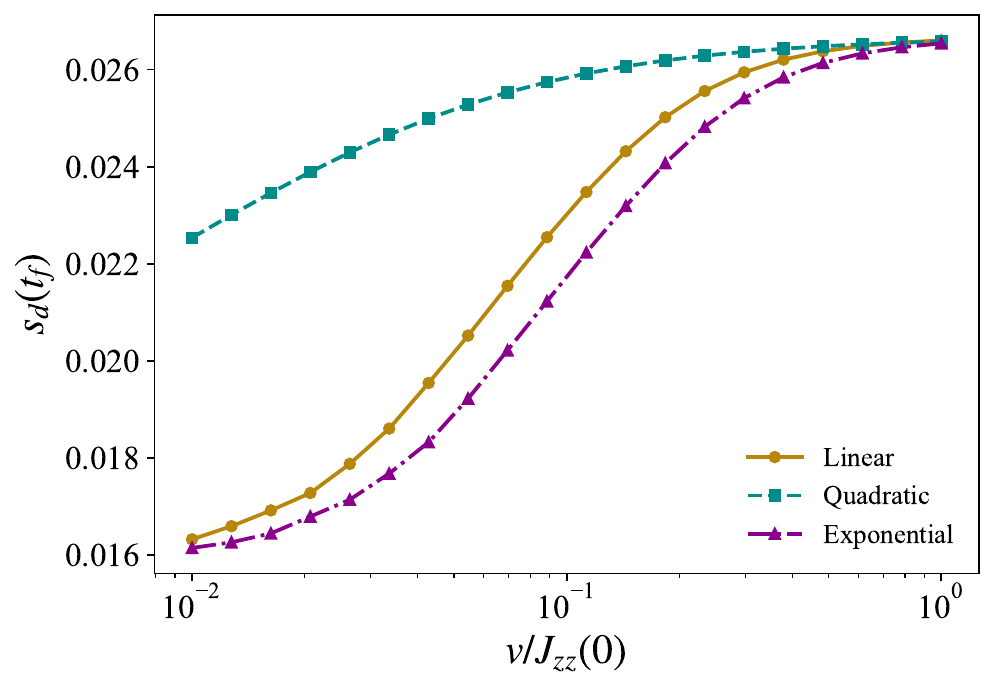}
	\end{minipage}
	\hfill
	\begin{minipage}[t]{0.48\textwidth}
		\includegraphics[width=\textwidth]{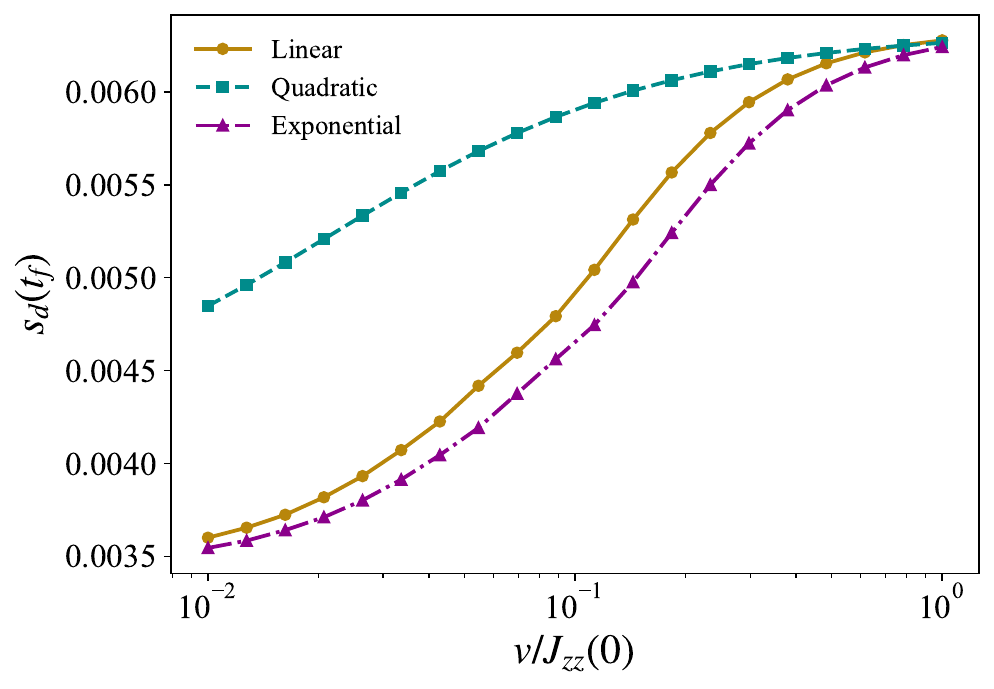}
	\end{minipage}
	
	\vspace{0.1cm}
	
	\begin{minipage}[t]{0.48\textwidth}
		\includegraphics[width=\textwidth]{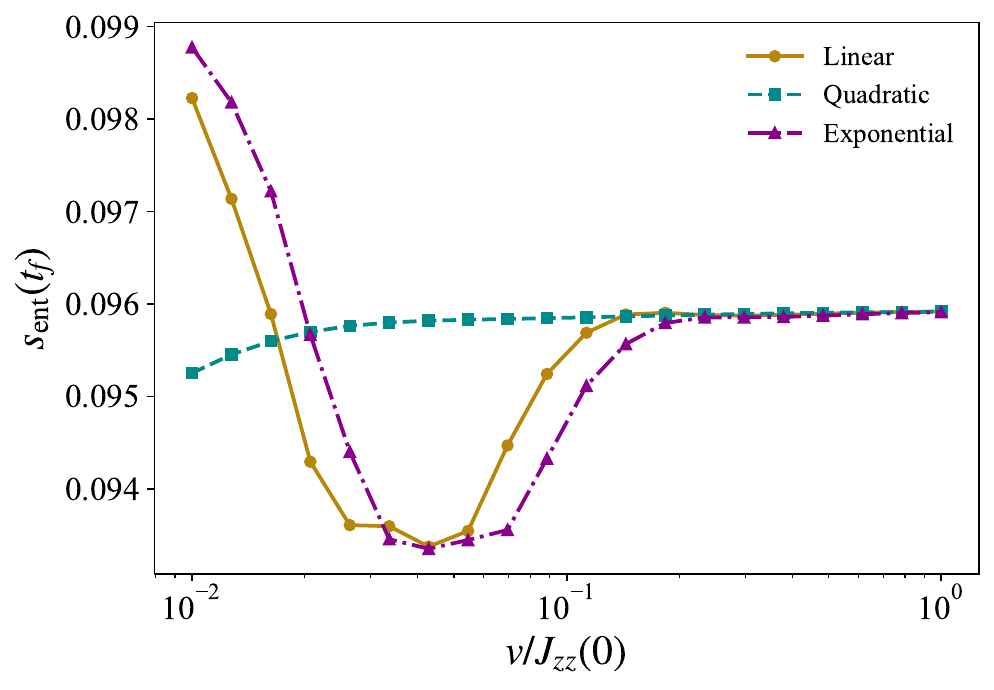}
	\end{minipage}
	\hfill
	\begin{minipage}[t]{0.48\textwidth}
		\includegraphics[width=\textwidth]{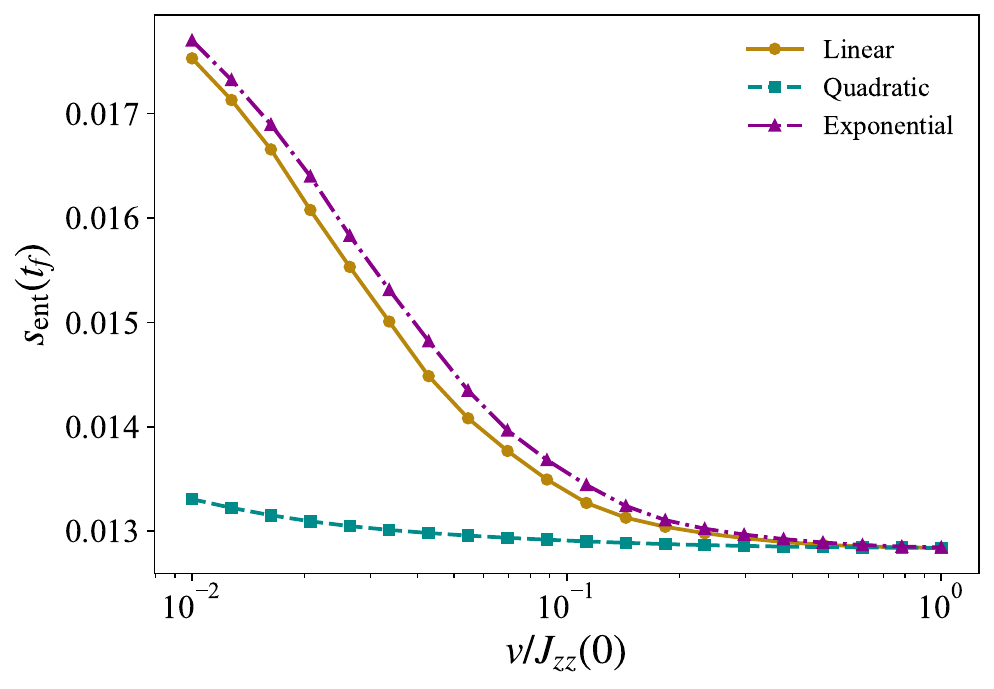}
	\end{minipage}
	\par\vspace{0.1cm}
	\centerline{\small (a) $L=10$}
	
	\vspace{0.4cm}
	
	\begin{minipage}[t]{0.48\textwidth}
		\includegraphics[width=\textwidth]{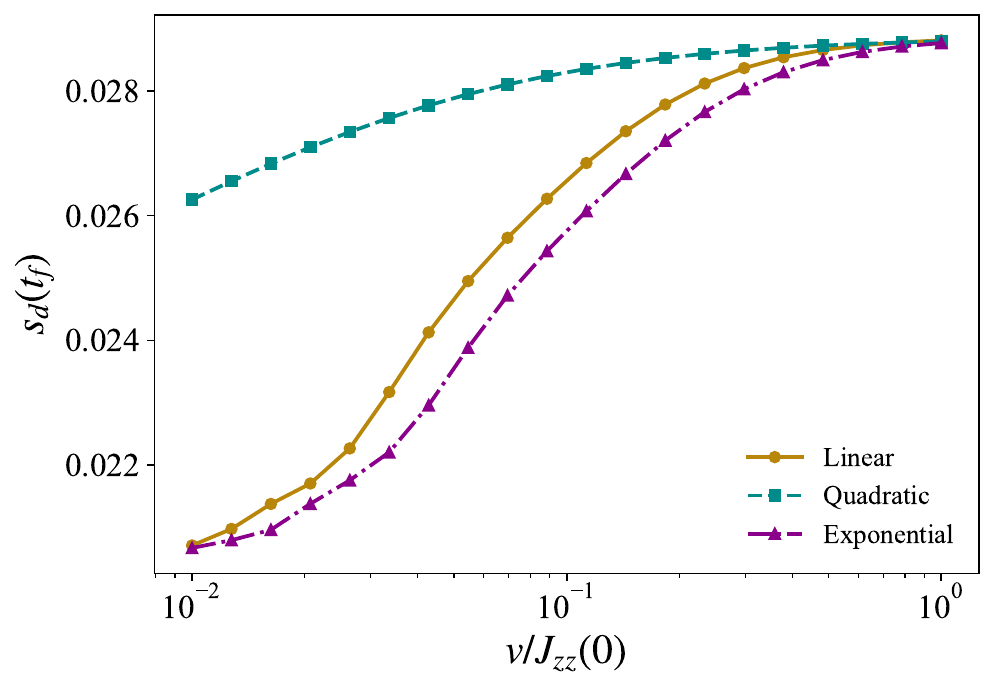}
	\end{minipage}
	\hfill
	\begin{minipage}[t]{0.48\textwidth}
		\includegraphics[width=\textwidth]{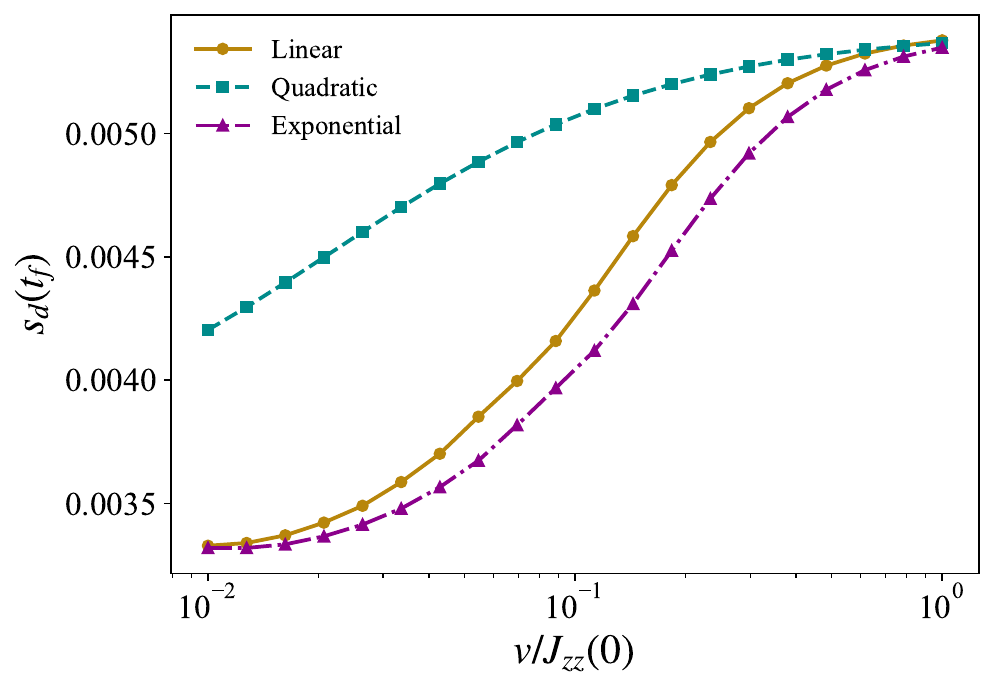}
	\end{minipage}
	
	\vspace{0.1cm}
	
	\begin{minipage}[t]{0.48\textwidth}
		\includegraphics[width=\textwidth]{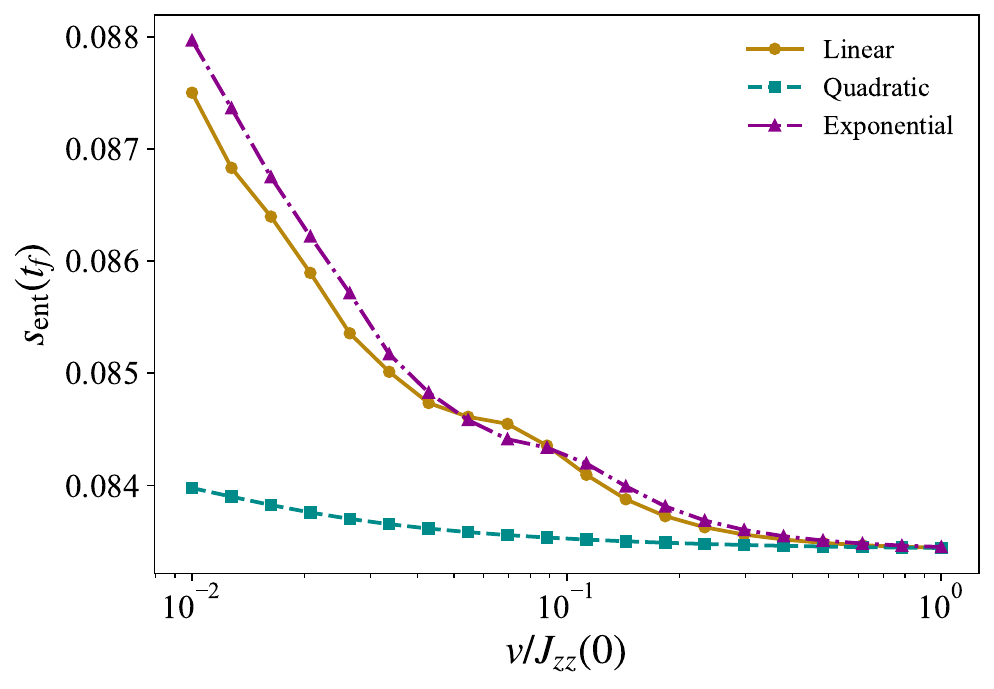}
	\end{minipage}
	\hfill
	\begin{minipage}[t]{0.48\textwidth}
		\includegraphics[width=\textwidth]{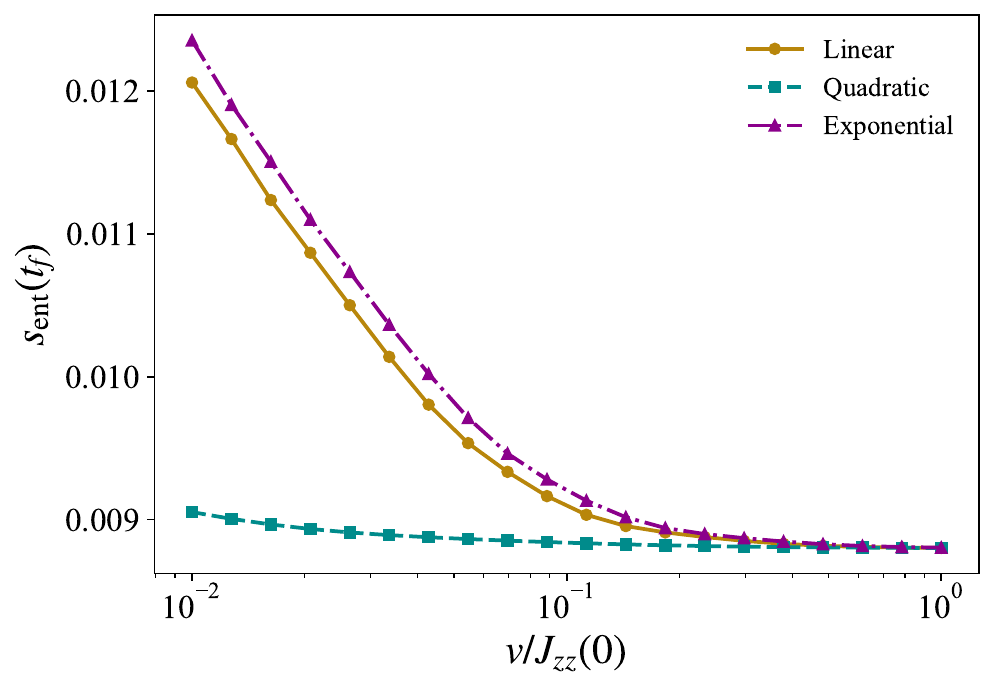}
	\end{minipage}
	\par\vspace{0.1cm}
	\centerline{\small (b) $L=12$}
	
\end{figure*}
\newpage

\begin{figure*}
	\centering
	\begin{minipage}[t]{0.48\textwidth}
		\includegraphics[width=\textwidth]{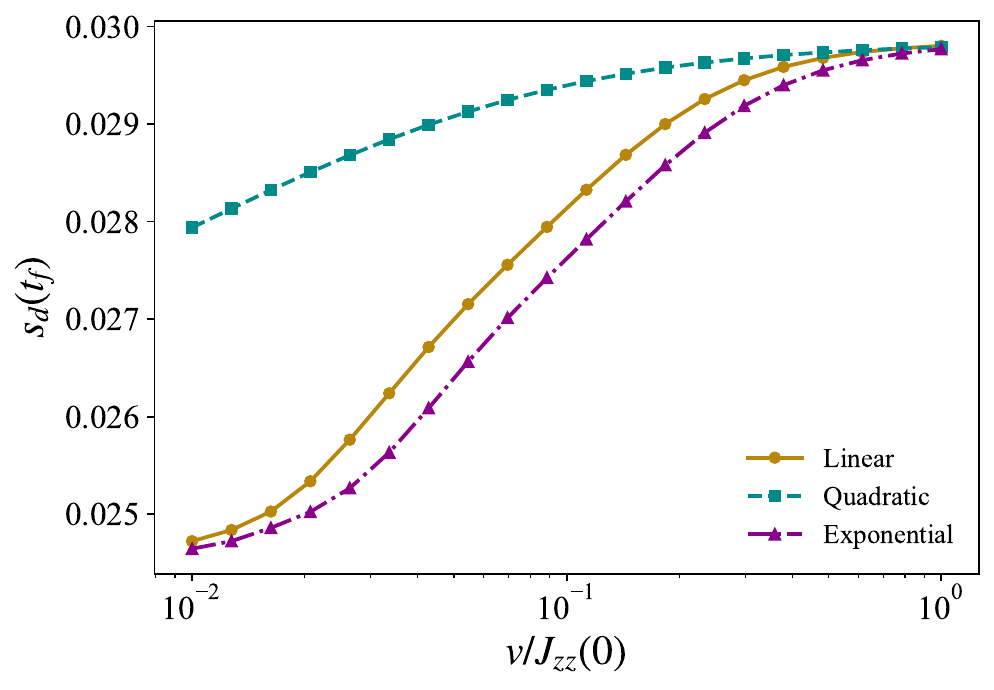}
	\end{minipage}
	\hfill
	\begin{minipage}[t]{0.48\textwidth}
		\includegraphics[width=\textwidth]{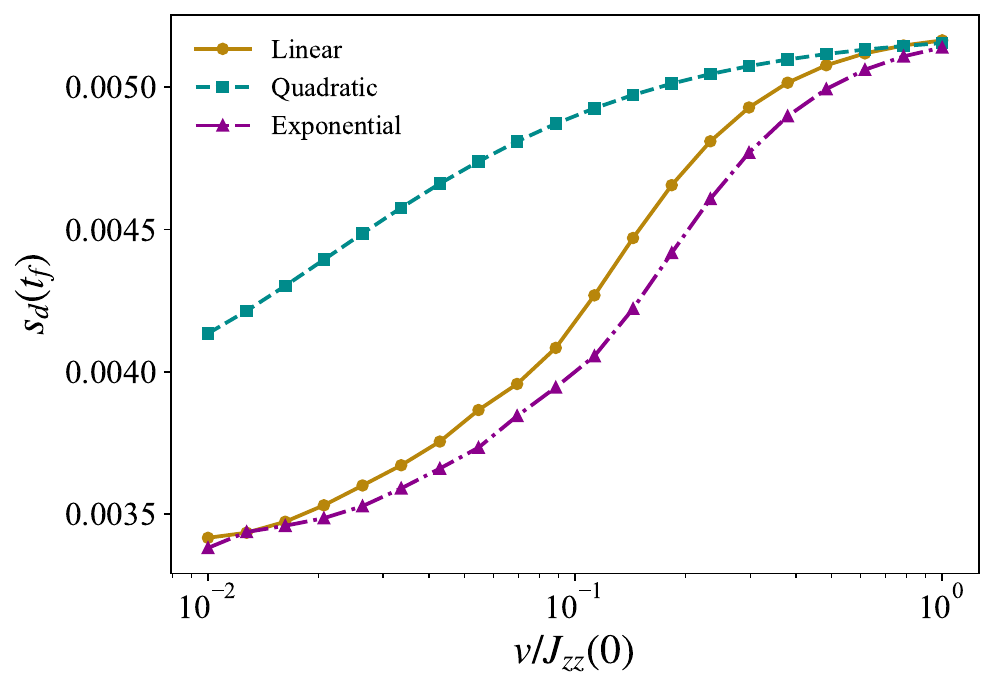}
	\end{minipage}
	
	\vspace{0.1cm}
	
	\begin{minipage}[t]{0.48\textwidth}
		\includegraphics[width=\textwidth]{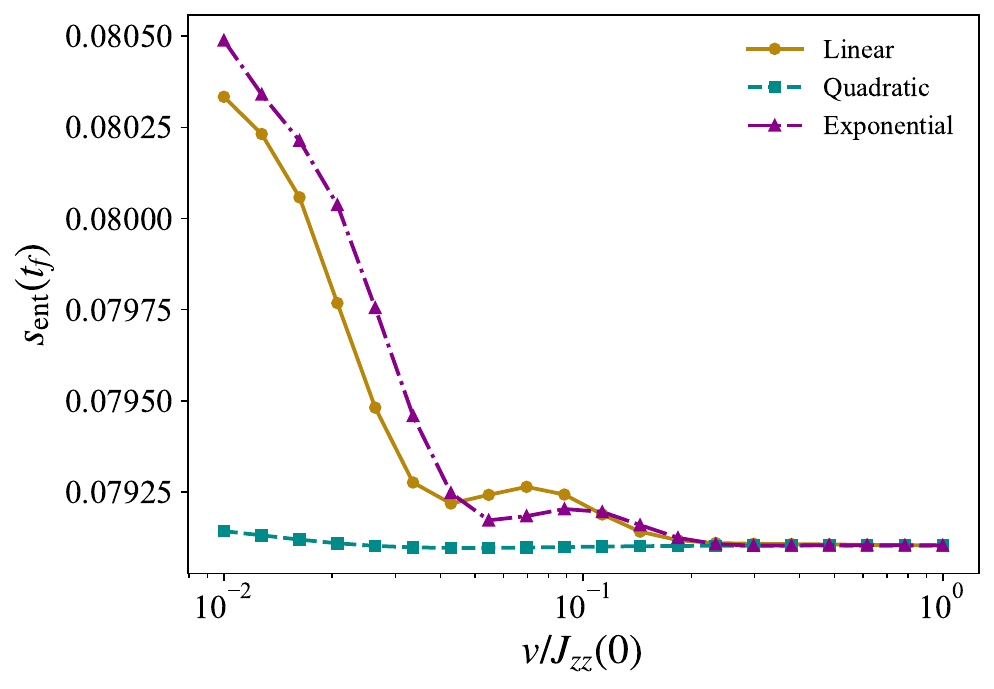}
	\end{minipage}
	\hfill
	\begin{minipage}[t]{0.48\textwidth}
		\includegraphics[width=\textwidth]{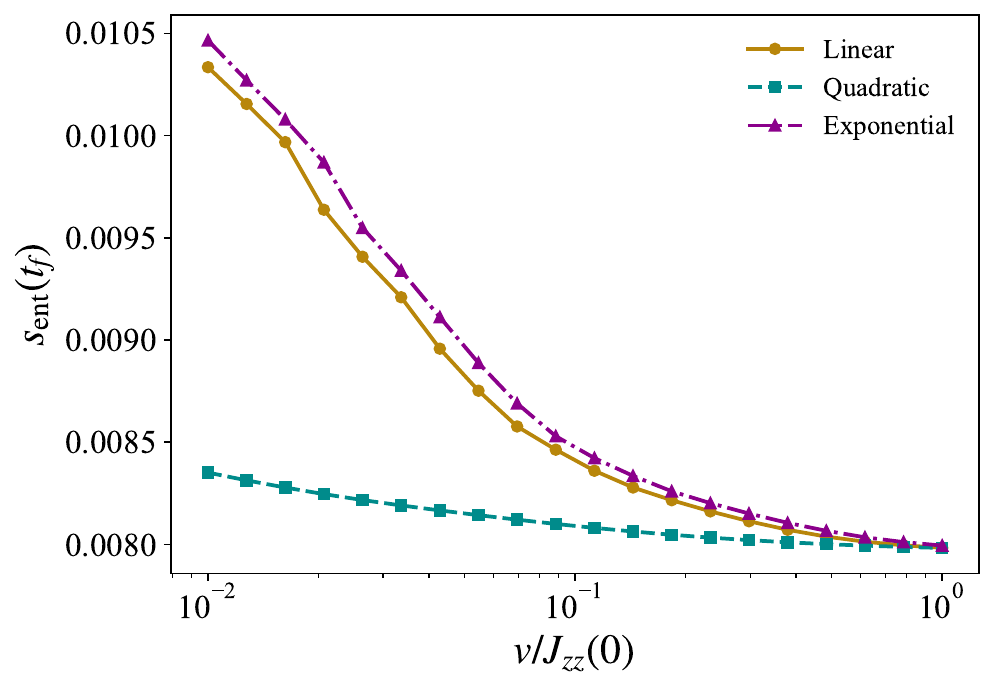}
	\end{minipage}
	\par\vspace{0.1cm}
	\centerline{\small (c) $L=14$}
	\caption{ Diagonal entropy density $s_d(t_f)$ and entanglement entropy density $s_{\mathrm{ent}}(t_f)$ for ETH and MBL phase. Panel (a)-(c) for $L=10$,  $L=12$ and $L=14$.}
	\label{fig:main_entropy_plots}
\end{figure*}

\subsection{Finite-Size Scaling and Protocol Robustness}

Figures~\ref{fig:main_entropy_plots}(a),
Figures~\ref{fig:main_entropy_plots}(b),
and Figures~\ref{fig:main_entropy_plots}(c) show the diagonal entropy density $s_d(t_f)$ together with the entanglement entropy density $s_{\mathrm{ent}}(t_f)$ that we obtain for linear, quadratic, and exponential ramps. This is done for sizes $L=10$, $12$, and $14$. If you look closely, this lets us do a direct check of finite-size effects and also how the ramp geometry shapes the nonequilibrium time evolution. Overall, the behavior looks like it barely cares about system size over the range we can actually access. More precisely, as $L$ increases, both scaling trend entropy measures remain unchanged. The absence of significant size-dependent rearrangements indicates that the observed dynamical behavior is not dominated by finite-size effects.

The ramp protocol primarily affects the magnitude and smoothness of the entropy response. Among the three protocols, the linear ramp gives the largest variation in both $s_d(t_f)$ and $s_{\mathrm{ent}}(t_f)$ , whereas the quadratic ramp ends up with a more tightly packed response. The exponential ramp instead leads to the smoothest-looking curves and the smallest spread range, which points to less sensitivity to finite-rate driving.
A simple interpretation is that the exponential protocol begins more gently, so it constraint the early-time accumulation of nonadiabatic excitations during that initial part of the ramp.
Even with those quantitative differences, the big picture remains pretty much the same across all $L$ values and all three drive profiles. Since the scaling trends remain intact, unchanged, this indicates that the dynamical signatures we extract from both diagonal and entanglement entropy are robust, not only against finite-size variation, but also against the exact temporal shape of the drive.

\section{Conclusion}
\label{conclusion}

We have studied the nonequilibrium dynamics of a disordered spin-$1/2$ XXZ chain, across the ETH--MBL crossover, while the interaction ramped is time-dependent. With exact diagonalization, we looked at the diagonal entropy density $s_d(t_f)$ and also the entanglement entropy density $s_{\mathrm{ent}}(t_f)$ when the ramp follows linear, quadratic, and exponential driving protocols. Our results show a pretty clear contrast between the ergodic and localized regimes. In the ETH phase, both entropy measures tend to grow as the ramp speed increases, suggesting that more excitation is generated and that the many-body states are mixed more efficiently. The MBL shows much weaker entropy production and does not care that much about driving rate, and it matches the idea that transport is suppressed and that correlation spreading is slowed down by localization.

The protocol matters for how large the response is, but the overall qualitative picture stays basically the same. The quadratic ramp gives the smallest entropy changes, while the linear and exponential ramps end up producing responses that are kind of comparable. Also, these behavior appears for every system size we checked, so it feels safe to say that the key conclusions are not just some finite-size effect. So in summary, diagonal entropy and entanglement entropy act like complementary probes for the ETH--MBL crossover when we drive at finite rates. The fact that their behavior stays stable against both the ramp shape and the system size underlines entropy production as a useful diagnostic tool for nonequilibrium dynamics in disordered interacting quantum systems, even if the details are slightly different depending on on how the control parameter is tuned in time.

\begin{acknowledgments}
	
V.A. acknowledges support from the Science and Engineering Research Board (SERB), Anusandhan–National Research Foundation (ANRF), and the Government of India, through Core Research Grant No. CRG/2023/001573. The computational resources provided by the PARAM Shavak (“Gryphon”) high-performance computing facility are also gratefully acknowledged.

\end{acknowledgments}

\section*{data availability}

The data that support the findings of this article are available upon reasonable request.

\bibliography{ref}

\end{document}